\documentclass[twocolumn]{aastex63}
\hypersetup{linkcolor=red,citecolor=blue,filecolor=cyan,urlcolor=blue}

\usepackage{amssymb}
\usepackage{graphicx}
\usepackage{natbib}
\usepackage[rightcaption]{sidecap}
\usepackage{txfonts}
\usepackage{units}
\usepackage{url}
\usepackage{pifont}
\usepackage[normalem]{ulem}

\definecolor{blue}{rgb}{0.00, 0.00, 1.00}
\definecolor{red}{rgb}{0.86, 0.08, 0.24}
\definecolor{orange}{rgb}{1.00, 0.55, 0.00}
\definecolor{darkblue}{rgb}{0.00, 0.00, 0.55}
\definecolor{green}{rgb}{0.00, 0.39, 0.00}
\definecolor{pink}{rgb}{1.000000,0.078431,0.576471}

\definecolor{myDarkRed}{rgb}{0.698, 0.094, 0.133}

\usepackage{ulem}
\definecolor{myBlue}{rgb}{.2,.7,.9}

\received{\today}
\revised{later}
\accepted{later}
\submitjournal{ApJ}

\shorttitle{Tracking downflows in an arch filament system}
\shortauthors{Gonz\'alez Manrique et al.}

\begin{document}

\title{Tracking downflows from the chromosphere to the photosphere in a solar arch filament system}

\correspondingauthor{Sergio Javier Gonz\'alez Manrique}
\email{smanrique@ta3.sk}

\author[0000-0002-6546-5955]{Sergio Javier Gonz\'alez Manrique}
\affil{Astronomical Institute, Slovak Academy of Sciences, 
05960 Tatransk\'{a} Lomnica, Slovak Republic}

\author[0000-0002-3242-1497]{Christoph Kuckein}
\affiliation{Leibniz-Institut f{\"u}r Astrophysik Potsdam (AIP),
An der Sternwarte 16, 14482 Potsdam, Germany}

\author{Adur Pastor Yabar}
\affiliation{Leibniz-Institut f{\"u}r Sonnenphysik (KIS), Sch\"oneckstra\ss{}e 6, 79104, Freiburg, Germany}

\author[0000-0002-9858-0490]{Andrea Diercke}
\affiliation{Leibniz-Institut f{\"u}r Astrophysik Potsdam (AIP),
An der Sternwarte 16, 14482 Potsdam, Germany}
\affiliation{Universit\"at Potsdam, Institut f\"ur Physik und Astronomie, 
Karl-Liebknecht-Str. 24-25, 14476 Potsdam, Germany}

\author{Manuel Collados}
\affiliation{Instituto de Astrof{\'i}sica de Canarias,
c/ V{\'i}a L{\'a}ctea s/n, 38205 La Laguna, Tenerife, Spain}

\author[0000-0002-0473-4103]{Peter G{\"o}m{\"o}ry}
\affiliation{Astronomical Institute, Slovak Academy of Sciences, 
05960 Tatransk\'{a} Lomnica, Slovak Republic}

\author[0000-0002-5606-0411]{Sihui Zhong}
\affiliation{CAS Key Laboratory of Solar Activity, National Astronomical Observatories,
Chinese Academy of Sciences, Beijing 100101, People's Republic of China}
\affiliation{University of Chinese Academy of Sciences, Beijing 100049, People's Republic of China}

\author[0000-0002-9534-1638]{Yijun Hou}
\affiliation{CAS Key Laboratory of Solar Activity, National Astronomical Observatories,
Chinese Academy of Sciences, Beijing 100101, People's Republic of China}
\affiliation{University of Chinese Academy of Sciences, Beijing 100049, People's Republic of China}

\author[0000-0002-7729-6415]{Carsten Denker}
\affiliation{Leibniz-Institut f{\"u}r Astrophysik Potsdam (AIP),
An der Sternwarte 16, 14482 Potsdam, Germany}



\begin{abstract}

We study the dynamics of plasma along the legs of an arch 
filament system (AFS) from the chromosphere to the photosphere, 
observed with high-cadence spectroscopic data
from two ground-based solar telescopes: the GREGOR 
telescope (Tenerife) using the GREGOR Infrarred Spectrograph (GRIS) 
in the \ion{He}{1} 10830~\AA\ range and the Swedish Solar Telescope 
(La Palma) using the CRisp Imaging Spectro-Polarimeter to observe
the \ion{Ca}{2} 8542~\AA\ and \ion{Fe}{1} 6173~\AA\ 
spectral lines. The temporal evolution of the draining of the plasma 
was followed along the legs of a single arch 
filament from the chromosphere to the photosphere. 
The average Doppler velocities inferred 
at the upper chromosphere from the \ion{He}{1} 10830~\AA\ triplet 
reach velocities up to 20\,--\,24~km~s$^{-1}$, in the lower chromosphere and 
upper photosphere the Doppler velocities reach up to 
11~km~s$^{-1}$ and 1.5~km~s$^{-1}$ in the case of the 
\ion{Ca}{2} 8542~\AA\ and \ion{Si}{1} 10827~\AA\ spectral lines, 
respectively. The evolution of the Doppler velocities 
at different layers of the solar atmosphere (chromosphere and upper photosphere) 
shows that they follow the same LOS velocity pattern, which confirm the observational evidence 
that the plasma drains towards the photosphere as proposed in models
of AFSs. The Doppler velocity maps inferred from the lower 
photospheric \ion{Ca}{1} 10839~\AA\ or \ion{Fe}{1} 6173~\AA\ spectral lines 
do not show the same LOS velocity pattern. Thus, there is 
no evidence that the plasma reaches the lower photosphere.
The observations and the nonlinear force-free field 
extrapolations demonstrate that the magnetic field loops of 
the AFS rise with time. We found flow asymmetries at different
footpoints of the AFS. The NLFFF values of the magnetic field strength
give us a clue to explain these flow asymmetries.

\end{abstract}




\section{Introduction}\label{SEC1}

Emerging flux regions (EFRs) are seen as magnetic concentrations 
in the photosphere of the Sun. From a theoretical point of view,
\citet{Parker1955} and \citet{Zwaan1987} proposed that EFRs are formed in the
convection zone and then emerge because of magnetic buoyancy (Parker
instability) to the solar surface. During the formation process of EFRs, 
merging and cancellation of different polarities occur, 
leading to various configurations of the magnetic field. Often, EFRs are
visible in the chromosphere in form of magnetic loops loaded with cool plasma
\citep{Solanki2003b}. They can be seen in the chromosphere as 
dark fibrils and they can reach up to the corona. Nowadays, 
we refer to them as an arch filament system 
\citep[AFS,][]{Bruzek1967, Bruzek1969} which connects two different 
polarities.

The AFSs are commonly observed in several spectral lines such as in the strong 
chromospheric absorption line H$\alpha$, or the line core of the 
\ion{Ca}{2} \,H \&\,K lines \citep[e.g.,][]{Bruzek1969, Su2018, Diercke2019}. 
AFSs can be observed in the 
\ion{He}{1} 10830~\AA\ triplet
\citep[e.g.,][]{Solanki2003b, Spadaro2004, Lagg2007, Xu2010, GonzalezManrique2018}.
This spectral line is formed in the upper chromosphere \citep{Avrett1994} and is
a very good candidate to observe chromospheric features and particularly AFSs.
Essentially, these structures develop with upflows in the 
midpoint of the loops and downflows at the footpoints 
\citep{Solanki2003b, GonzalezManrique2018}.
The upflows can reach velocities up to 20~km~s$^{-1}$ and the downflows 
at the footpoints are observed typically in a range between 30\,--\,50~km~s$^{-1}$
\citep[see, e.g.,][]{Solanki2003b, Balthasar2016, GonzalezManrique2017b,Zhong2019}.
Supersonic velocities at this chromospheric heights are considered above
$v$ $>$ 10\,km\,s$^{-1}$ \citep{AznarCuadrado2005}.
These high velocities translate into two components of the 
\ion{He}{1} 10830~\AA\ triplet, typically known as slow and fast components
\citep{Lagg2004}. The fast component reaches supersonic velocities and at some 
height generates a shock because of the transition from 
lower densities (corona) into higher densities (chromosphere and below). 
Hence, the \ion{He}{1} profiles are seen slightly in emission 
\citep{Lagg2007}. On the contrary, \citet{Xu2010} did not find any evidence 
for shocks in the \ion{He}{1} triplet and, therefore, proposed that the shock 
occurs below the formation height of \ion{He}{1}.


\begin{figure}[t]
\includegraphics[width=\columnwidth]{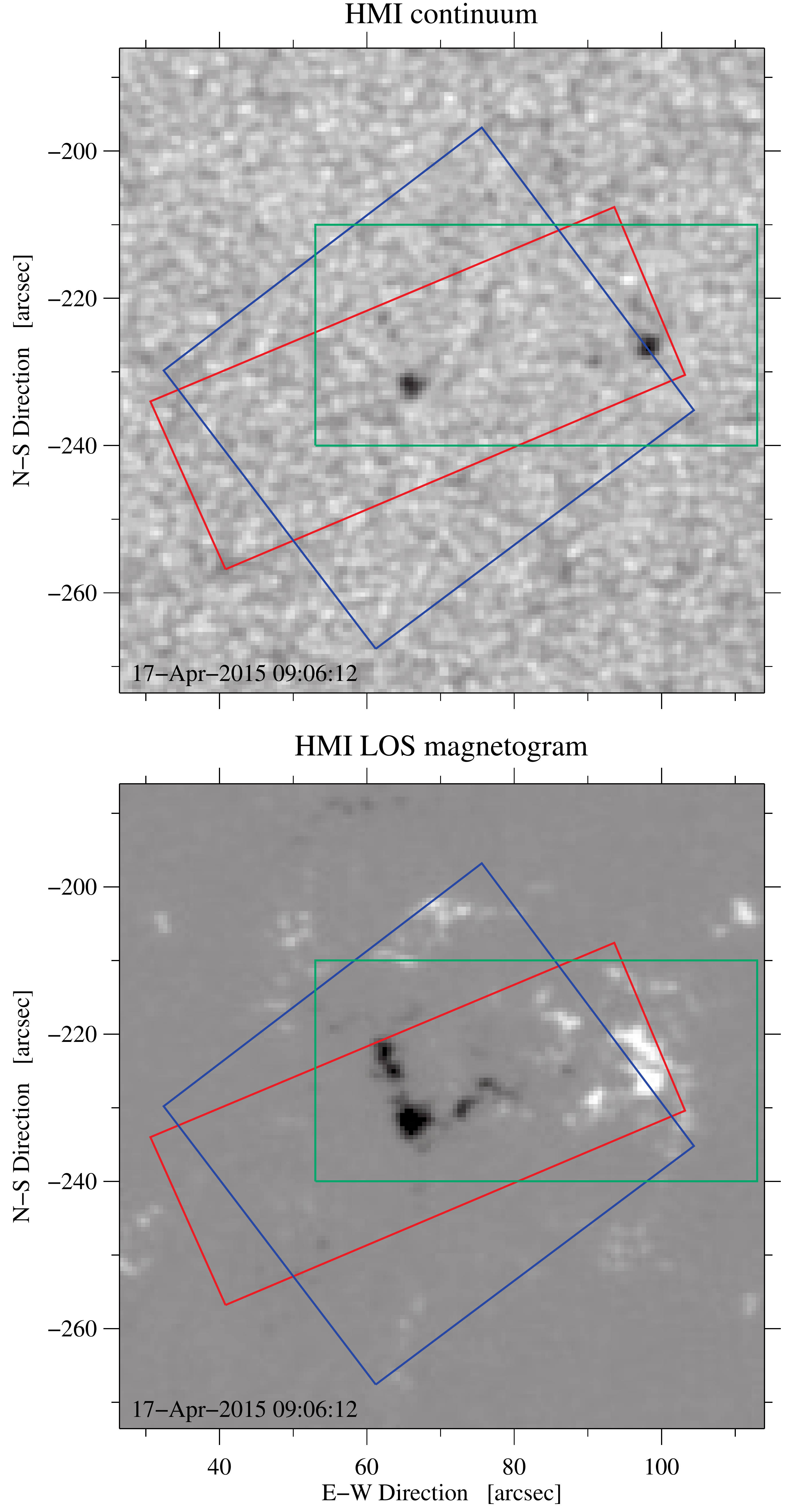}
\caption{Overview of the EFR at 09:06~UT on 2015~April~17: 
       SDO HMI continuum image (\textit{top}) and HMI magnetogram
       (\textit{bottom}). We depict the the FOV of GRIS (red), 
       CRISP (blue), and of the NLFFF results in Fig.~\ref{FIG04} 
       (green). }
\label{FIG01}
\end{figure}

This study is the continuation of
\citet{GonzalezManrique2017, GonzalezManrique2018}, who studied the 
evolution of an AFS in \ion{He}{1}. The goal of the present study 
is to follow the evolution of the plasma flows across several heights 
at the footpoints of an AFS.




%
%

\section{Observations and data reduction}\label{SEC2}

A small EFR, containing two pores with opposite polarities and 
an associated AFS in the chromosphere, was observed on 2015 April~17.
The region of interest (ROI) is located at heliographic coordinates 
S19$^{\circ}$ and W4$^{\circ}$ ($\mu \equiv \cos\theta = 0.97$). Two instruments placed at two 
telescopes were involved in this coordinated observing campaign: (1) the 
GREGOR Infrared Spectrograph \citep[GRIS,][]{Collados2012}
located at the 1.5-meter GREGOR solar telescope \citep{Schmidt2012}
at Observatorio del  Teide, Tenerife, Spain and (2) the CRisp Imaging
Spectro-Polarimeter \citep[CRISP,][]{Scharmer2008a} 
located at the Swedish Solar Telescope \citep[SST,][]{Scharmer2003a} 
at Observatorio Roque de los Muchachos, La Palma, Spain. The
overlap of the two field-of-views (FOV) of the instruments 
is shown in Fig.~\ref{FIG01}. The FOVs are aligned with a continuum 
image of the Helioseismic and Magnetic Imager 
\citep[HMI,][]{Scherrer2012, Schou2012} on board of the 
Solar Dynamics Observatory \citep[SDO,][]{Pesnell2012}.

We applied the standard data reduction to the spatio-spectral data
cubes of the very fast spectroscopic mode from GRIS \citep{GonzalezManrique2016, GonzalezManrique2018}. 
The wavelength calibration includes corrections of the solar 
gravity redshift and orbital-motion \citep[see appendices A and B in][]{Kuckein2012b}. 
Since telluric lines are present in our spectral range, the Doppler
velocities computed in this study were retrieved from an absolute-scale
wavelength calibrated array. The spectral region observed with GRIS
comprises the photospheric \ion{Ca}{1} 10839~\AA\ and \ion{Si}{1} 10827~\AA\ lines,
as well as the chromospheric \ion{He}{1} 10830~\AA\ triplet among others. 
\citet{GonzalezManrique2017,GonzalezManrique2018} studied the same data set of 
GRIS in the very fast spectroscopic mode in the \ion{He}{1} 10830~\AA\ spectral line. 
This study builds on the results of the previously mentioned papers.

Five time-series were recorded with CRISP each consisting of ten data sets with
full-Stokes measurements in the photospheric \ion{Fe}{1} 6173~\AA\ line and in
the chromospheric \ion{Ca}{2} 8542~\AA\ line. The observations cover the
evolution of the region between 08:47~UT and 9:20~UT with a FOV of $54\arcsec
\times 54\arcsec$. The spectral sampling of the \ion{Fe}{1} line consist of 
19 wavelength positions with an equidistant step of 25~m\AA\ (spectral range
$-225$~m\AA\ to $+225$~m\AA\ with respect to the central
wavelength and a continuum position at $+525$~m\AA). The exposure  
time amounts to 33~ms for a single image (12 accumulations, 33~ms each).

The spectral sampling of the chromospheric \ion{Ca}{2} line comprises 
21 wavelength positions. The exposure time amounts 
also to 33~ms for a single image (6 accumulations, 33~ms each).
Sequentially observing both lines yields a total cadence of about 50~s. 
The CRISPRED data pipeline \citep{delaCruzRodriguez2015} was used for CRISP data,
carrying out dark, flat-field, demodulation, prefilter 
corrections among others. The images were restored
with Multi-Object Multi-Frame Blind Deconvolution 
\citep[MOMFBD,][]{vanNoort2005} method, based on the 
algorithm by \citet{Loefdahl2002}. 

\begin{figure*}
\centering
\includegraphics[width=0.8\textwidth]{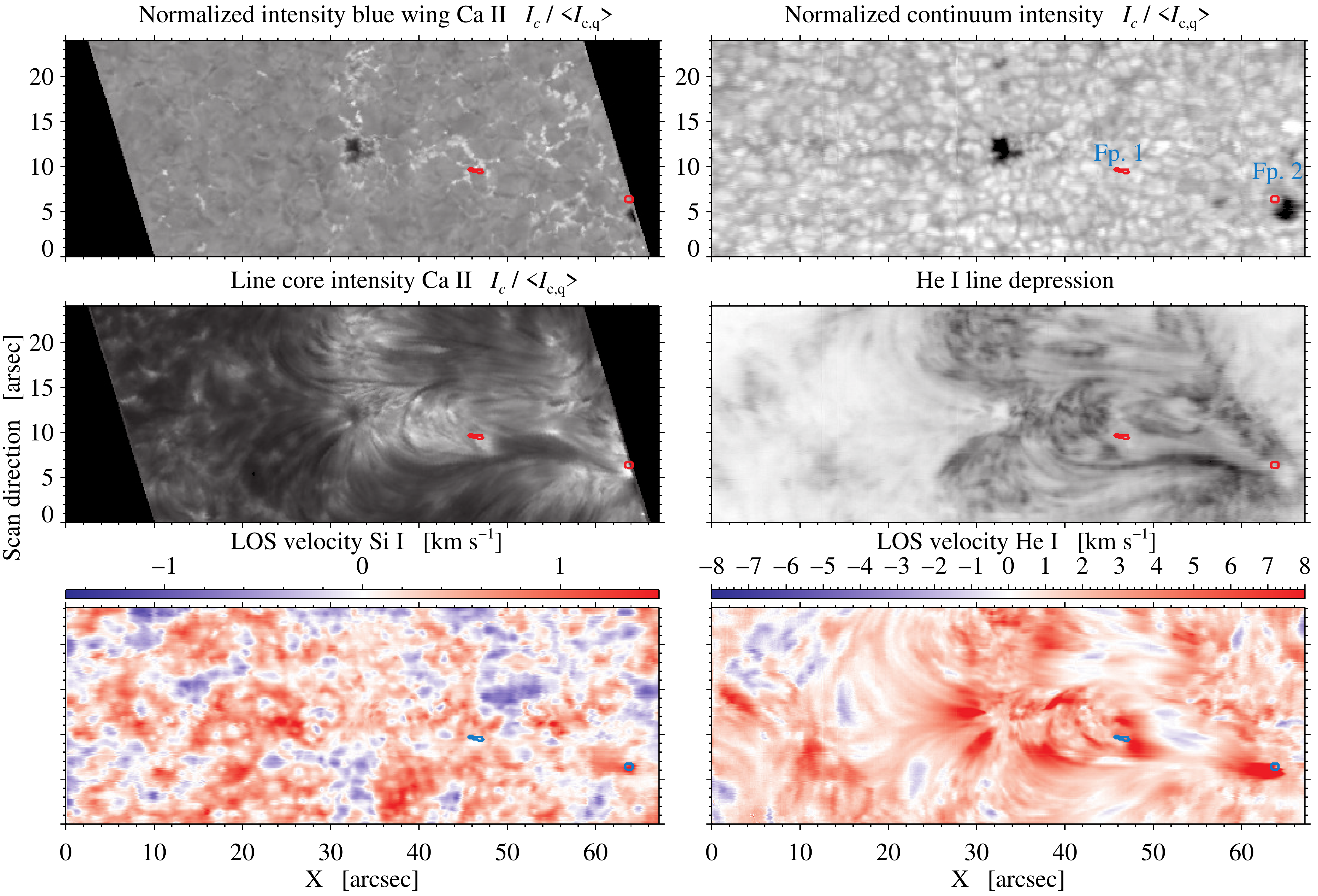}
\caption{Restored CRISP images (\textit{top and middle left}) at 09:05:27 UT 
        and slit-reconstructed GRIS images (\textit{top and middle right}) 
        at 09:05:54 UT on 2015~April~17 of the EFR: blue wing of the chromospheric
        \ion{Ca}{2} line (\textit{top left}), continuum intensity observed
        with GRIS (\textit{top right}), line core intensity of the
        chromospheric \ion{Ca}{2} line (\textit{middle left}), 
        \ion{He}{1} line depression (\textit{middle right}), 
        \ion{Si}{1} Doppler velocity (\textit{bottom left}), and 
        \ion{He}{1} Doppler velocity (\textit{bottom left}). 
        The red/blue contours depict the location of strong downflows at two
        different footpoints (Fp1 and Fp2). Black areas in the CRISP 
        images indicate that there is no overlapping data with GRIS.
 } 
\label{FIG02}
\end{figure*}

At around 09:05~UT, strong downflow velocities occur near one pore. 
The $50^\mathrm{th}$ scan at 09:05:54~UT (see Fig.~\ref{FIG02}) 
was selected as a reference for the GRIS data because of 
the good seeing conditions. 
The corresponding data set of CRISP was taken at 
09:05:27~UT, despite only fair seeing conditions prevailed at SST.

The HMI data were compensated for differential rotation 
with respect to the central meridian. The reference 
image was taken at 00:00:57~UT on 2015~April~17, 
considering that the position of the EFR was exactly 
at the central meridian.

%
%

\section{Data analysis}\label{SEC4}


A previous analysis of this data set was carried out for the Doppler 
velocities of the \ion{He}{1} 10830~\AA\ triplet by \citet[][]{GonzalezManrique2018}. 
This triplet consists of one blue and two blended red components. A small 
proportion of spectral profiles contains apparent signatures 
of ``dual flows'' \citep{Schmidt2000b}, which split the component 
into a slow and fast component. The single flow profiles of
the red component of the \ion{He}{1} triplet were fitted with a 
single Lorentzian. Furthermore, the dual-flow profiles were fitted with a 
double-Lorentzian profile. Details of the procedure on how to fit the two parts 
of the red component is explained in \citet{GonzalezManrique2016, GonzalezManrique2018}.


The \ion{Si}{1} 10827~\AA, the \ion{Ca}{1} 10839~\AA, and 
the \ion{Fe}{1} 6173~\AA\ spectral lines were fitted with a Gaussian 
using the Levenberg-Marquardt least-squares minimization as 
implemented in the MPFIT IDL software package \citep{Markwardt2009}, 
to infer the respective LOS velocities at the core. The wavelength references for 
the LOS velocities were set to the laboratory wavelengths 10827.09~\AA, 
10838.97~\AA, and 6173.33~\AA, respectively, which were taken from
the National Institute of Standards and Technology (NIST)\footnote{www.nist.gov} 
database.


The information encoded in the chromospheric \ion{Ca}{2} 8542~\AA\ spectral 
line was analyzed using the non-LTE inversion code NICOLE 
\citep{SocasNavarro2015}. The inversion process is complex 
and time-consuming for non-LTE lines. Hence, we concentrated only 
on a small region-of-interest (ROI) of 40 pixels (Fp.~1 in 
Fig.~\ref{FIG02}). This region was selected because we found strong
chromospheric \ion{He}{1} downflows during 30 minutes
(almost one hour in the case of the Fp.~2 with 36 pixels, not
inverted with NICOLE). We investigated also Fp.~2 because the
\ion{He}{1} Doppler velocities are higher than at Fp.~1 and the 
probability that the plasma reaches the photosphere is even higher
at Fp.~2 compared to Fp.~1. The areas selected 
are represented by a red/blue contour in Fig.~\ref{FIG02}.
The areas also depict the two regions with 
higher frequency of occurrence of 
\ion{He}{1} dual-flow profiles during the observing period with 
GRIS \citep[Fig.~5 in ][]{GonzalezManrique2018}. In some 
pixels we find persistently dual-flow profiles in 60 out of 64 \ion{He}{1} maps.
Since the aim is to retrieve the LOS
velocities, we focused only on the inversion of the Stokes-$I$
profiles inside this region. For the inversions, we took into
account the isotopic splitting of the \ion{Ca}{2} NIR line
\citep[][]{Leenaarts2014} and used the FALC model
\citep[][]{Fontenla1993} as an initial estimate for the atmosphere. 


To investigate the evolution of the three-dimensional (3D) 
magnetic topology of the AFS, we perform nonlinear 
force-free field (NLFFF) extrapolations by using the
``weighted optimization'' method 
\citep{Wiegelmann2004, Wiegelmann2012}. The boundary condition for the 
NLFFF extrapolation is given by an HMI photospheric vector magnetogram 
with an image scale of 0.5\arcsec~pixel$^{-1}$ which was
preprocessed by a procedure developed by \citet{Wiegelmann2006} to satisfy 
the force-free condition. The NLFFF extrapolations are performed 
within a box of $224\,\times\,104\,\times\,128$ uniformly 
spaced grid points (about 81 $\,\times\,38\,\times\,46$\,Mm$^{3}$).

%
%

\section{Results}\label{SEC5}


In this study we investigate the height dependence 
of the draining flows along the arch filaments across different layers of the solar
atmosphere, from the upper chromosphere down to the photosphere.  

Following the plasma requires carefully selected spectral 
lines observed simultaneously, which form at different heights 
of the solar atmosphere and cover a large range of heights. In this case,
two chromospheric and three photospheric spectral lines were 
used, combining two different ground-based telescopes. 

\begin{figure*}
\includegraphics[width=0.8\textwidth]{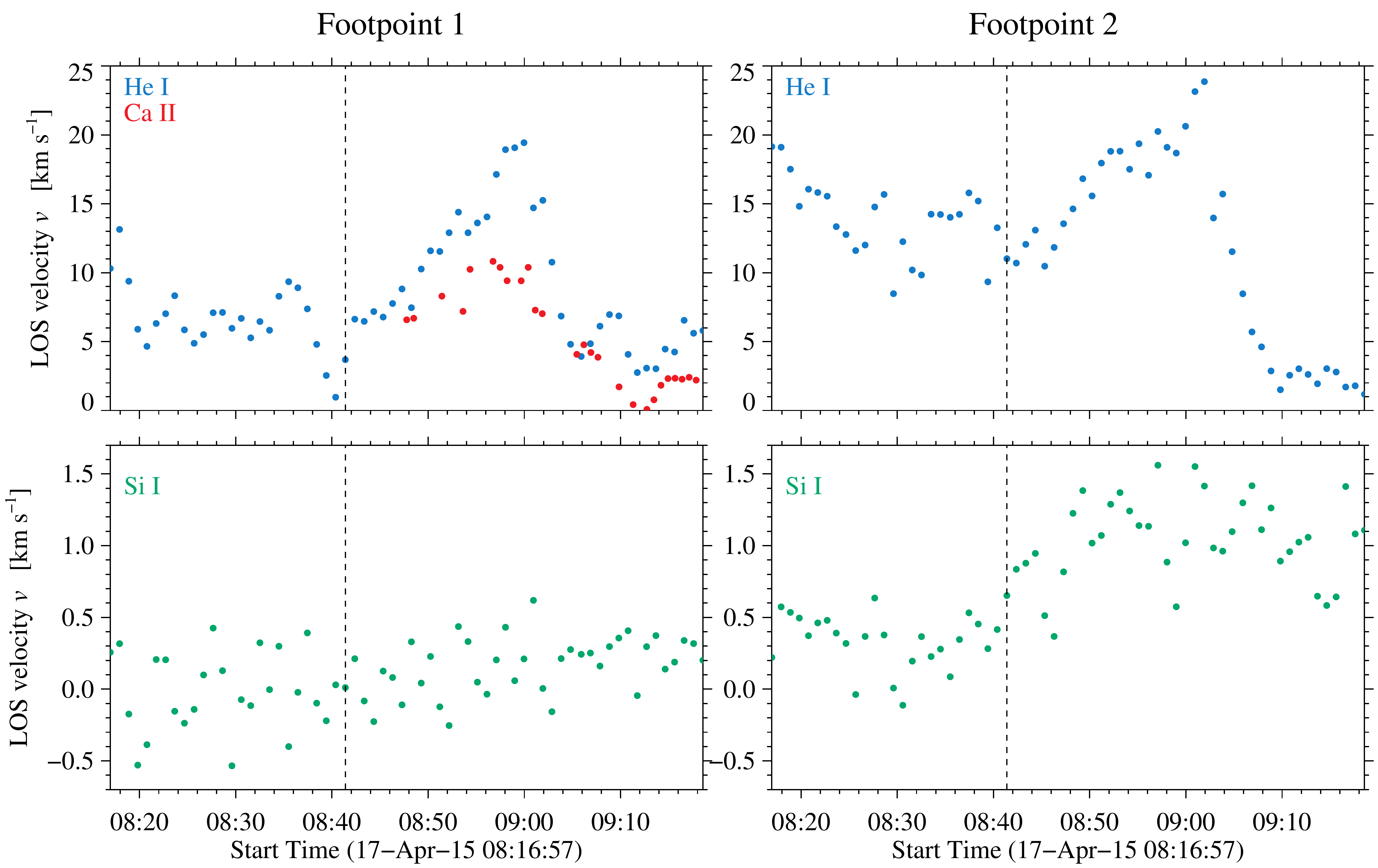}
\centering
\caption{Temporal evolution of the average Doppler velocities
        of Fp.~1 (\textit{left}) and 
        Fp.~2 (\textit{right}) as observed in the
        \ion{He}{1} triplet red component (\textit{blue}), 
        the \ion{Ca}{2} spectral line (\textit{red}), and the
        \ion{Si}{1} spectral line (see Fig.~\ref{FIG02}). The dashed vertical
        lines mark the time when the Doppler velocities suddenly increase. 
} 
\label{FIG03}
\end{figure*}

We selected the two footpoints of a single arch filament (see Fp.~1 and 
Fp.~2 in Fig.~\ref{FIG02}). We computed the average Doppler velocities
within the area of the contours of Fp.~1 and Fp.~2 in every map available with
both instruments. In Fig.~\ref{FIG03}, we show the temporal evolution of the
Doppler velocities for the \ion{He}{1} red component, \ion{Ca}{2}, and \ion{Si}{1}. 

The temporal evolution of the average Doppler velocities based on the \ion{He}{1}
triplet's red component and the \ion{Si}{1} spectral line of Fp.~2 is 
represented by blue and green bullets in the top right and bottom right panels of
Fig.~\ref{FIG03}. Practically the entire CRISP data do not include the Fp.~2 area.
Hence, it was impossible to calculate the evolution of the Doppler velocities 
at Fp.~2 based on the \ion{Ca}{2} and \ion{Fe}{1} spectral lines.

In both footpoints Fp.~1 and Fp.~2, the temporal evolution of the
Doppler velocities based on the \ion{He}{1} red component 
follows a similar pattern, showing peaks of high velocities at around 09:00~UT.
The flows represent the evolution taking place in the upper layers of the
chromosphere. Between 08:17~UT and 08:41~UT of the time-series observed with GRIS
the average Doppler velocities varied in the range of
1\,--\,13~km~s$^{-1}$ for Fp.~1 whereas Fp.~2 fluctuated between
7\,--\,19~km~s$^{-1}$. Around 08:41~UT 
(dashed line in Fig.~\ref{FIG03}), the averaged 
Doppler velocities rapidly increased up to 20 and 24~km~s$^{-1}$ at 
Fp.~1 (09:00~UT) and Fp.~2 (09:02~UT), respectively. At the end of 
the time-series, starting at 09:08~UT the Doppler velocities strongly
dropped to 0\,--\,7~km~s$^{-1}$ for Fp.~1 and 2\,--\,4km~s$^{-1}$ for Fp.~2, respectively.

The temporal evolution between 08:47~UT and  09:20~UT of the mean 
Doppler velocities based on the \ion{Ca}{2} is delineated by red bullets 
(in Fig.~\ref{FIG03}). The Doppler velocities at Fp.~1 were computed as averaged values 
within the range of $\log \tau \in [-2.4, -3.0]$ (red/blue contour in Fig.~\ref{FIG02}), 
corresponding to the upper photosphere. This range fits within the values of 
the computed response functions (RFs) by \citet{QuinteroNoda2016} and 
\citet{Kuckein2017b} for \ion{Ca}{2} 8542~\AA. The plot exhibits the same peak 
at around 09:00~UT as the \ion{He}{1} Doppler shifts.
The average Doppler velocities promptly increased up to 11~km~s$^{-1}$ 
and then rapidly dropped to almost 0~km~s$^{-1}$.
Thus, we assume that the plasma moves along the leg of 
the arch filament at Fp.~1, from the upper to the lower chromosphere/upper photosphere
with a clear deceleration.

The line core of the \ion{Si}{1} spectral line is formed 
in the upper photosphere \citep[e.g.,][]{Bard2008,Shchukina2017,Felipe2016}. 
However, when computing the Doppler shifts with a Gaussian
fit, we retrieve the average shift of the line. 
The LOS velocity evolution at Fp.~1 
does not show any clear sign that the plasma reaches
the upper photosphere at the \ion{Si}{1} height formation
(green bullets in Fig.~\ref{FIG02}). 
What we observe is likely the convection pattern with Doppler
velocities varying between $\pm$0.5~km~s$^{-1}$. Conversely, the 
velocity evolution at Fp.~2 shows an increase of the
velocities measured in \ion{Si}{1} co-temporal to the one
exhibited by the \ion{He}{1} triplet in the upper 
chromosphere. At the beginning of the time-series the 
velocity reaches up to 0.5~km~s$^{-1}$. The velocity then
rapidly increases at about the same time as 
the \ion{He}{1} triplet (dashed line in Fig.~\ref{FIG03}) 
reaching velocities up to 1.5~km~s$^{-1}$. Finally, 
the LOS velocities drop a few minutes later than in the upper 
chromosphere, at around 09:10~UT. The average Doppler velocities computed for the photospheric
\ion{Fe}{1} and \ion{Ca}{1} photospheric spectral lines 
do not show any correlation with the strong downflows 
seen in the upper chromosphere.


\begin{figure*}
\includegraphics[width=0.8\textwidth]{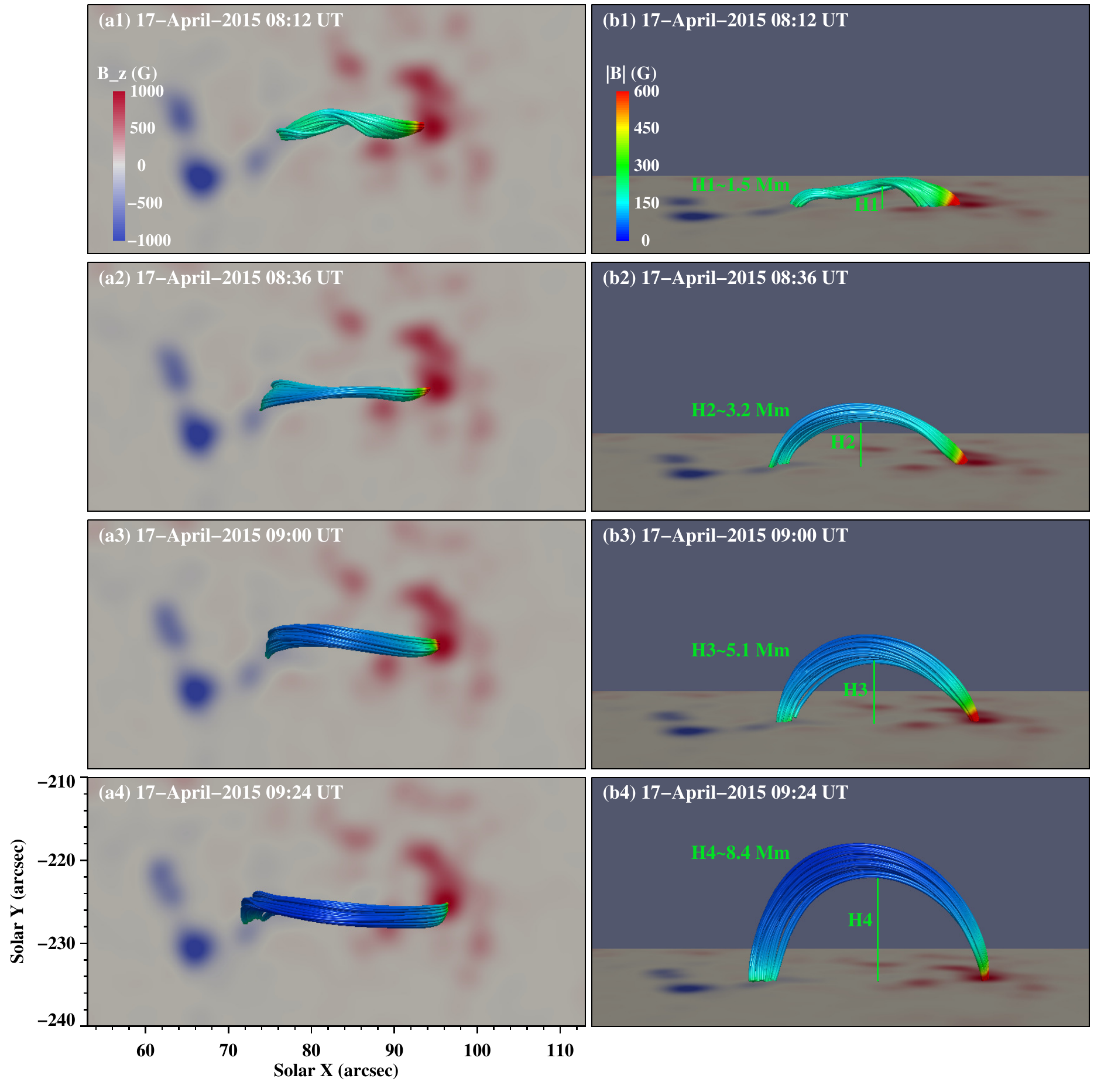}
\centering
\caption{Magnetic structures revealed by NLFFF extrapolations
        for the arch filament studied in Fig.~\ref{FIG03} connecting
        the footpoints Fp.~1 and Fp.~2 (Fig.~\ref{FIG01}). Four points in time
        were selected to compute the extrapolations: before
        the observations (\textit{top}), first indications of the rising loop (\textit{second from top}),
        during the time of the strongest Doppler velocities were detected as shown in
        Fig.~\ref{FIG03} (\textit{third}), and right after
        the observations (\textit{bottom}). The photospheric vertical 
        magnetograms ($B_z$) are displayed as background. The left 
        panels depict the top view of the structures.
        The right panels show the side view. The colored loops
        illustrate the AFS magnetic fields. The different colors of the field lines
        depict the different values of the local magnetic field strength where red is the highest value.
} 
\label{FIG04}
\end{figure*}

The NLFFF extrapolations were modeled for four different times 
to associate the flows to the magnetic field topology  
(08:12~UT, 08:36~UT, 09:00~UT, and 09:24~UT). The time range covers our ground-based observing time.
Figure~\ref{FIG04} shows different perspectives of the extrapolation results. 
The colored loops represent how the estimated magnetic 
fields strength of the AFS evolve with time.
Close to the beginning of the observations with GRIS, the extrapolations exhibit
slightly twisted magnetic loops with maximum height of about 1.5~Mm. 
Approximately 24 and 48~minutes later, the magnetic loops are no longer twisted and
the height increased up to 3.2 and 5.1~Mm respectively. At the end of the time-series,
the height of the magnetic loop continued to increase up to 8.4~Mm. This confirms the
hypothesis of \citet{GonzalezManrique2018} concerning the evolution
of an arch filament (see their Fig.~15). The scenario presented by these authors
suggests step by step how the plasma rises from the photosphere 
to the chromosphere and even the corona. The sketch only represents 
the evolution as seen in the \ion{He}{1}. In the first
panels (Fig.~15) they described how the plasma rises from the photosphere to the chromosphere
and the velocities at the loop tops reach their maximum 
upflow velocities.
After a few minutes the material starts to drain along the legs towards the
photosphere. This plasma accelerates reaching supersonic velocities. 
As seen in the last panels (Fig.~15), the arch filament continues rising to the 
transition region or even the corona. The downflows observed at the
footpoints progressively decrease until they completely vanish. 
The NLFFF extrapolations and the LOS velocities
presented in this study corroborate this scenario.

In Fig.~\ref{FIG04}, the field lines of the emerging magnetic loops were colored depending 
on the values of local magnetic field strength. This clearly
shows strong magnetic fields at the footpoints and weaker fields at the loop tops. 
Interestingly, Fp.~2 exhibits higher values than the values computed at 
Fp.~1. This can be an important aspect to discern the downflow asymmetries between
both footpoints (see Sect.~\ref{SEC6}).

\newpage

%
%

\section{Discussion and Conclusions}\label{SEC6}

%
%


We scrutinized how the plasma flows along the footpoints of 
an AFS connecting two different polarities through different layers of the solar
atmosphere. In both footpoints the dynamic shows 
similar LOS velocity pattern at different atmospheric layers, the upper chromosphere
and the upper photosphere. This similar behavior of the plasma flows at 
different layers confirms that the plasma reached lower layers of the solar
atmosphere coming from the upper chromosphere. 


To confirm the plasma reaching the upper photosphere at Fp.~2, we 
used \ion{He}{1} and \ion{Si}{1} (GRIS), while at Fp.~1, 
we also used \ion{Ca}{2} (CRISP). As we demonstrate in Fig.~\ref{FIG03}, we do not 
detect the plasma reaching  the upper photosphere with the Doppler 
velocities inferred from the \ion{Si}{1} spectral line at 
Fp.~1 while at Fp.~2 they are well detected. We expected to observe
the same behaviour at Fp.~1 because the same LOS velocity pattern is identified 
in the upper photosphere with the \ion{Ca}{2} spectral line. 
However, the \ion{Si}{1} line is formed deeper in the atmosphere than the
\ion{Ca}{2} line and the fast flows have not reached the layers to 
which the \ion{Si}{1} line is sensitive to. On the contrary,  
the plasma reaches deeper layers of the solar atmosphere at Fp.~2, compared to
Fp.~1. We did not find any emission signatures in \ion{He}{1}. The shocks
manifest themselves in emission in this triplet. Therefore, there is no evidence
for emission. Consequently, we cannot confirm that the plasma decelerates because of shocks produced
below the \ion{He}{1} height formation. A shock scenario cannot explain the downflow asymmetries in our case.  
We propose a different scenario based on the NLFFF extrapolations. The extrapolations 
(Fig.~\ref{FIG04}) do not show any obvious asymmetry 
of the rising loop, e.g., twist or an inclination angle. 
Furthermore, the extrapolations present different values of
the magnetic field strength along the legs of the loops. 
The leg at Fp.~2 clearly
has stronger magnetic fields $B$ (up to 600~G) than the leg
at Fp.~1. Interestingly, the field lines at Fp.~2 are more compact or concentrated 
than those found at Fp.~1, which suggests a smaller cross-sectional area at Fp.~2. 
Hence, the magnetic field 
strength $B$ together with the cross-sectional areas $S$ are the key to understand 
the downflow asymmetries, as explained below.

We assume that the plasma moves from the chromosphere to the 
photosphere along a flux tube. The cross-sectional area $S$ of the 
tube is decreasing in the downward direction, and the plasma becomes
denser. The plasma motion is governed by the equations of continuity, motion, and energy. 
The conservation of mass is given by the differential form of the continuity
equation
\begin{equation}
\frac{\partial \rho}{\partial t} + \nabla (\rho v) = 0,
\label{eq3}
\end{equation}
where $\rho$ is the mass density and the Lagrangian derivative $d / dt 
\equiv \partial / \partial t + v \cdot \nabla$ was applied. If we also assume that 
the fluid is stationary $\partial \rho / \partial t = 0$, the dynamic
equilibrium is given by
\begin{equation}
\nabla (\rho v) = 0.
\label{eq4}
\end{equation}
If in addition the fluid moves along the field lines, then the equation 
simplifies to
\begin{equation}
\rho v S = c_1,
\label{eq5}
\end{equation}
where $S$ is the area enclosing the field lines, and $c_1$ is a
constant.
Similarly, one of Maxwell's equations simplifies
\begin{equation}
\nabla B = 0,
\label{eq6}
\end{equation}
where $B$ is the magnetic field. If the magnetic flux is conserved then
\begin{equation}
BS = c_2,
\label{eq7}
\end{equation}
where $S$ is again the area enclosing the field lines, and $c_2$ is another
constant.


Following the aforementioned equations, a plausible explanation for the flow 
asymmetries is that the values of the magnetic 
field $B$ are higher at Fp.~2 compared to Fp.~1. Stronger magnetic fields mean tighter
field lines along this leg (the lower the height is, the stronger is the magnetic field $B$
and the cross-sectional area $S$ decreases). Consequently, the lower the height the smaller 
the cross section at Fp.~2 compared to Fp.~1. If we take into account the equations above (Eq.~\ref{eq5} and \ref{eq7}), 
together with the magnetic field values obtained from the extrapolations 
($\sim$600~G at Fp.~2 versus $\sim$300~G at Fp.~1 at the level of the photosphere) and assuming that the
density $\rho$ is similar at both footpoints, the velocities necessarily need to be higher
(about two times higher) at Fp.~2 compared to Fp.~1.
Since the field strength is stronger at Fp.~2, owing to the conservation of the flux across
the flux tube the cross-sectional area $S$ needs to be smaller and hence the LOS velocities are higher
at Fp.~2 than at Fp.~1. The photosperic \ion{Si}{1} velocities measured
at Fp.~1 are around $\pm$0.6~km~s$^{-1}$ and between 0~km~s$^{-1}$ and 1.5~km~s$^{-1}$ at Fp.~2 (see Fig.~\ref{FIG03}).
These values are in line with the proposed scenario. This explains 
the downflow asymmetries between both footpoints. 
In addition, it elucidates why the plasma slows down sooner along the leg of the AFS at Fp.~1 
and does not reach the upper photosphere, as inferred from the \ion{Si}{1} Doppler shifts,
whereas at Fp.~2 we have evidence that the plasma reaches this height (as inferred from the 
\ion{Ca}{2} inversions).


Downflows in arch filaments were explained by \citet{Chou1993}
as the emergence of a flux tube into the solar atmosphere. 
\citet{Lagg2007} proposed that the plasma carried by the rising loops
drains to lower layers of the solar atmosphere along its legs because of 
the effect of gravity and the concurrent needs for vertical 
hydrostatic equilibrium and horizontal pressure balance.
\citet{GonzalezManrique2018} proposed that
the arch filament carries plasma during the rise of the arch filament
from the photosphere to the corona (see their sketch in Fig.~15). 
The authors suggested that after a certain 
time after the AFS starts rising, the plasma drains towards the photosphere
along their legs reaching chromospheric supersonic velocities.
Based on NLFFF extrapolations, we demonstrate that 
the magnetic field loops of the arch filament studied here rise with time
(in about one hour) from 1.5~Mm up to 8.4~Mm, confirming the
hypothesis of \citet{GonzalezManrique2018} concerning the evolution
of an arch filament. Interestingly, the loop reaches the largest height at 09:24~UT
but the plasma flows already decline around 09:10~UT (with the maximum around 09:00~UT). We propose
that the magnetic loop continues to rise but the plasma inside of the loop 
is evacuated before the magnetic loop reaches the corona. We observe that the
\ion{He}{1} absorption at the loop vanishes at the end of our observations (See the available
movie in \citet{GonzalezManrique2018}). Furthermore, we do not observe dual-flows 
anymore at the footpoints at the end of the observations (after 09:10~UT).
Consequently, it is not possible to observe high velocities either because they do 
not exist any longer or because we cannot see them in the available spectral lines of this study.

\acknowledgments

The 1.5-m GREGOR solar telescope was built by a 
German  consortium  under  the  leadership  of  the  Leibniz-Institut  f{\"u}r  
Sonnenphysik  in  Freiburg  with  the  AIP,  the
Institut f{\"u}r Astrophysik G{\"o}ttingen, and the Max-Planck-Institut f{\"u}r 
Sonnensystemforschung in G{\"o}ttingen as partners, and with contributions by the Instituto de
Astrof{\'i}sica de Canarias and the Astronomical Institute of the Academy of 
Sciences of the Czech Republic. SDO HMI data are provided by the Joint Science
Operations Center – Science Data Processing. This work is part of a 
collaboration between the Astronomical Institute of 
Slovak Academy of Sciences and AIP 
supported by the German Academic Exchange Service (DAAD), with funds from 
the German Federal Ministry of Education \& Research and Slovak Academy of 
Sciences, under project No. 57449420. SJGM and PG acknowledge the support 
of the project VEGA 2/0048/20. SJGM also is grateful for the support of the 
Stefan Schwarz grant of the  Slovak Academy of Sciences and the 
support by the Erasmus+ programme of the European Union under grant number 
2017-1-CZ01-KA203-035562 during his 2019 stay at the Instituto de 
Astrof{\'i}sica de Canarias. APY is grateful 
for the support by the German DFG project number 321818926.
MC acknowledges financial support from the Spanish
Ministerio de Ciencia, Innovaci{\'o}n y Universidades through project
PGC2018-102108-B-I00 and FEDER funds. Hou Y.J. is supported by the National 
Natural Science Foundations of China (11903050 and 11790304). 
Funding from the Horizon 2020 projects SOLARNET (No 824135) and ESCAPE
(No 824064) is greatly acknowledged.
We would like to thank Dr. H{\'e}ctor Socas  for  enlightening 
discussions about NICOLE inversions. We would like to thank
the referee who providing helpful comments and guidance,
improving the structure and contents of this manuscript.


\bibliography{apj-jour,sergio}

\begin{thebibliography}{}
\expandafter\ifx\csname natexlab\endcsname\relax\def\natexlab#1{#1}\fi
\providecommand{\url}[1]{\href{#1}{#1}}
\providecommand{\dodoi}[1]{doi:~\href{http://doi.org/#1}{\nolinkurl{#1}}}
\providecommand{\doeprint}[1]{\href{http://ascl.net/#1}{\nolinkurl{http://ascl.net/#1}}}
\providecommand{\doarXiv}[1]{\href{https://arxiv.org/abs/#1}{\nolinkurl{https://arxiv.org/abs/#1}}}

\bibitem[{{Avrett} {et~al.}(1994){Avrett}, {Fontenla}, \&
  {Loeser}}]{Avrett1994}
{Avrett}, E.~H., {Fontenla}, J.~M., \& {Loeser}, R. 1994, in IAU Symp., Vol.
  154, Infrared Solar Physics, ed. D.~M. {Rabin}, J.~T. {Jefferies}, \&
  C.~{Lindsey}, 35

\bibitem[{{Aznar Cuadrado} {et~al.}(2005){Aznar Cuadrado}, {Solanki}, \&
  {Lagg}}]{AznarCuadrado2005}
{Aznar Cuadrado}, R., {Solanki}, S.~K., \& {Lagg}, A. 2005, in ESA Spec. Publ.,
  Vol. 596, Chromospheric and Coronal Magnetic Fields, ed. D.~E. {Innes},
  A.~{Lagg}, \& S.~K. {Solanki}, 49.1

\bibitem[{{Balthasar} {et~al.}(2016){Balthasar}, {G{\"o}m{\"o}ry},
  {Gonz{\'a}lez Manrique}, {Kuckein}, {Kavka}, {Ku{\v c}era}, {Schwartz},
  {Va{\v s}kov{\'a}}, {Berkefeld}, {Collados Vera}, {Denker}, {Feller},
  {Hofmann}, {Lagg}, {Nicklas}, {Orozco Su{\'a}rez}, {Pastor Yabar}, {Rezaei},
  {Schlichenmaier}, {Schmidt}, {Schmidt}, {Sigwarth}, {Sobotka}, {Solanki},
  {Soltau}, {Staude}, {Strassmeier}, {Volkmer}, {von der L{\"u}he}, \&
  {Waldmann}}]{Balthasar2016}
{Balthasar}, H., {G{\"o}m{\"o}ry}, P., {Gonz{\'a}lez Manrique}, S.~J., {et~al.}
  2016, Astron. Nachr., 337, 1050, \dodoi{10.1002/asna.201612432}

\bibitem[{{Bard} \& {Carlsson}(2008)}]{Bard2008}
{Bard}, S., \& {Carlsson}, M. 2008, \apj, 682, 1376, \dodoi{10.1086/589910}

\bibitem[{{Bruzek}(1967)}]{Bruzek1967}
{Bruzek}, A. 1967, \solphys, 2, 451, \dodoi{10.1007/BF00146493}

\bibitem[{{Bruzek}(1969)}]{Bruzek1969}
---. 1969, \solphys, 8, 29, \dodoi{10.1007/BF00150655}

\bibitem[{{Chou}(1993)}]{Chou1993}
{Chou}, D.~Y. 1993, in ASP Conf. Ser., Vol.~46, IAU Colloq. 141: The Magnetic
  and Velocity Fields of Solar Active Regions, ed. H.~{Zirin}, G.~{Ai}, \&
  H.~{Wang}, 471--478

\bibitem[{{Collados} {et~al.}(2012){Collados}, {L{\'o}pez}, {P{\'a}ez},
  {Hern{\'a}ndez}, {Reyes}, {Calcines}, {Ballesteros}, {D{\'{\i}}az}, {Denker},
  {Lagg}, {Schlichenmaier}, {Schmidt}, {Solanki}, {Strassmeier}, {von der
  L{\"u}he}, \& {Volkmer}}]{Collados2012}
{Collados}, M., {L{\'o}pez}, R., {P{\'a}ez}, E., {et~al.} 2012, Astron. Nachr.,
  333, 872, \dodoi{10.1002/asna.201211738}

\bibitem[{{de la Cruz Rodr{\'{\i}}guez} {et~al.}(2015){de la Cruz
  Rodr{\'{\i}}guez}, {L{\"o}fdahl}, {S{\"u}tterlin}, {Hillberg}, \& {Rouppe van
  der Voort}}]{delaCruzRodriguez2015}
{de la Cruz Rodr{\'{\i}}guez}, J., {L{\"o}fdahl}, M.~G., {S{\"u}tterlin}, P.,
  {Hillberg}, T., \& {Rouppe van der Voort}, L. 2015, \aap, 573, A40,
  \dodoi{10.1051/0004-6361/201424319}

\bibitem[{{Diercke} {et~al.}(2019){Diercke}, {Kuckein}, \&
  {Denker}}]{Diercke2019}
{Diercke}, A., {Kuckein}, C., \& {Denker}, C. 2019, \aap, 629, A48,
  \dodoi{10.1051/0004-6361/201935583}

\bibitem[{{Felipe} {et~al.}(2016){Felipe}, {Collados}, {Khomenko}, {Kuckein},
  {Asensio Ramos}, {Balthasar}, {Berkefeld}, {Denker}, {Feller}, {Franz},
  {Hofmann}, {Joshi}, {Kiess}, {Lagg}, {Nicklas}, {Orozco Su{\'a}rez}, {Pastor
  Yabar}, {Rezaei}, {Schlichenmaier}, {Schmidt}, {Schmidt}, {Sigwarth},
  {Sobotka}, {Solanki}, {Soltau}, {Staude}, {Strassmeier}, {Volkmer}, {von der
  L{\"u}he}, \& {Waldmann}}]{Felipe2016}
{Felipe}, T., {Collados}, M., {Khomenko}, E., {et~al.} 2016, Astronomy and
  Astrophysics, 596, A59, \dodoi{10.1051/0004-6361/201629586}

\bibitem[{{Fontenla} {et~al.}(1993){Fontenla}, {Avrett}, \&
  {Loeser}}]{Fontenla1993}
{Fontenla}, J.~M., {Avrett}, E.~H., \& {Loeser}, R. 1993, \apj, 406, 319

\bibitem[{{Gonz{\'a}lez Manrique} {et~al.}(2017{\natexlab{a}}){Gonz{\'a}lez
  Manrique}, {Bello Gonz{\'a}lez}, \& {Denker}}]{GonzalezManrique2017b}
{Gonz{\'a}lez Manrique}, S.~J., {Bello Gonz{\'a}lez}, N., \& {Denker}, C.
  2017{\natexlab{a}}, \aap, 600, A38, \dodoi{10.1051/0004-6361/201527880}

\bibitem[{{Gonz{\'a}lez Manrique} {et~al.}(2016){Gonz{\'a}lez Manrique},
  {Kuckein}, {Pastor Yabar}, {Collados}, {Denker}, {Fischer}, {G{\"o}m{\"o}ry},
  {Diercke}, {Bello Gonz{\'a}lez}, {Schlichenmaier}, {Balthasar}, {Berkefeld},
  {Feller}, {Hoch}, {Hofmann}, {Kneer}, {Lagg}, {Nicklas}, {Orozco Su{\'a}rez},
  {Schmidt}, {Schmidt}, {Sigwarth}, {Sobotka}, {Solanki}, {Soltau}, {Staude},
  {Strassmeier}, {Verma}, {Volkmer}, {von der L{\"u}he}, \&
  {Waldmann}}]{GonzalezManrique2016}
{Gonz{\'a}lez Manrique}, S.~J., {Kuckein}, C., {Pastor Yabar}, A., {et~al.}
  2016, Astron. Nachr., 337, 1057

\bibitem[{{Gonz{\'a}lez Manrique} {et~al.}(2017{\natexlab{b}}){Gonz{\'a}lez
  Manrique}, {Denker}, {Kuckein}, {Pastor Yabar}, {Collados}, {Verma},
  {Balthasar}, {Diercke}, {Fischer}, {G{\"o}m{\"o}ry}, {Bello Gonz{\'a}lez},
  {Schlichenmaier}, {Cubas Armas}, {Berkefeld}, {Feller}, {Hoch}, {Hofmann},
  {Lagg}, {Nicklas}, {Orozco Su{\'a}rez}, {Schmidt}, {Schmidt}, {Sigwarth},
  {Sobotka}, {Solanki}, {Soltau}, {Staude}, {Strassmeier}, {Volkmer}, {von der
  L{\"u}he}, \& {Waldmann}}]{GonzalezManrique2017}
{Gonz{\'a}lez Manrique}, S.~J., {Denker}, C., {Kuckein}, C., {et~al.}
  2017{\natexlab{b}}, in IAU Symposium, Vol. 327, IAU Symposium, ed. S.~{Vargas
  Dom{\'{\i}}nguez}, A.~G. {Kosovichev}, P.~{Antolin}, \& L.~{Harra}, 28--33,
  \dodoi{10.1017/S1743921317000278}

\bibitem[{{Gonz{\'a}lez Manrique} {et~al.}(2018){Gonz{\'a}lez Manrique},
  {Kuckein}, {Collados}, {Denker}, {Solanki}, {G{\"o}m{\"o}ry}, {Verma},
  {Balthasar}, {Lagg}, \& {Diercke}}]{GonzalezManrique2018}
{Gonz{\'a}lez Manrique}, S.~J., {Kuckein}, C., {Collados}, M., {et~al.} 2018,
  \aap, 617, A55, \dodoi{10.1051/0004-6361/201832684}

\bibitem[{{Kuckein} {et~al.}(2012){Kuckein}, {Mart{\'{\i}}nez Pillet}, \&
  {Centeno}}]{Kuckein2012b}
{Kuckein}, C., {Mart{\'{\i}}nez Pillet}, V., \& {Centeno}, R. 2012, \aap, 542,
  A112, \dodoi{10.1051/0004-6361/201218887}

\bibitem[{{Kuckein} {et~al.}(2017){Kuckein}, {Diercke}, {Gonz{\'a}lez
  Manrique}, {Verma}, {L{\"o}hner-B{\"o}ttcher}, {Socas-Navarro}, {Balthasar},
  {Sobotka}, \& {Denker}}]{Kuckein2017b}
{Kuckein}, C., {Diercke}, A., {Gonz{\'a}lez Manrique}, S.~J., {et~al.} 2017,
  \aap, 608, A117, \dodoi{10.1051/0004-6361/201731319}

\bibitem[{{Lagg} {et~al.}(2004){Lagg}, {Woch}, {Krupp}, \&
  {Solanki}}]{Lagg2004}
{Lagg}, A., {Woch}, J., {Krupp}, N., \& {Solanki}, S.~K. 2004, \aap, 414, 1109,
  \dodoi{10.1051/0004-6361:20031643}

\bibitem[{{Lagg} {et~al.}(2007){Lagg}, {Woch}, {Solanki}, \&
  {Krupp}}]{Lagg2007}
{Lagg}, A., {Woch}, J., {Solanki}, S.~K., \& {Krupp}, N. 2007, \aap, 462, 1147,
  \dodoi{10.1051/0004-6361:20054700}

\bibitem[{{Leenaarts} {et~al.}(2014){Leenaarts}, {de la Cruz Rodr{\'{\i}}guez},
  {Kochukhov}, \& {Carlsson}}]{Leenaarts2014}
{Leenaarts}, J., {de la Cruz Rodr{\'{\i}}guez}, J., {Kochukhov}, O., \&
  {Carlsson}, M. 2014, \apjl, 784, L17

\bibitem[{{L{\"o}fdahl}(2002)}]{Loefdahl2002}
{L{\"o}fdahl}, M.~G. 2002, in \procspie, Vol. 4792, Image Reconstruction from
  Incomplete Data, ed. P.~J. {Bones}, M.~A. {Fiddy}, \& R.~P. {Millane},
  146--155, \dodoi{10.1117/12.451791}

\bibitem[{{Markwardt}(2009)}]{Markwardt2009}
{Markwardt}, C.~B. 2009, in ASP Conf.\ Ser., Vol. 411, Astronomical Data
  Analysis Software and Systems XVIII, ed. D.~A. {Bohlender}, D.~{Durand}, \&
  P.~{Dowler}, 251--254

\bibitem[{{Parker}(1955)}]{Parker1955}
{Parker}, E.~N. 1955, \apj, 121, 491, \dodoi{10.1086/146010}

\bibitem[{{Pesnell} {et~al.}(2012){Pesnell}, {Thompson}, \&
  {Chamberlin}}]{Pesnell2012}
{Pesnell}, W.~D., {Thompson}, B.~J., \& {Chamberlin}, P.~C. 2012, \solphys,
  275, 3

\bibitem[{{Quintero Noda} {et~al.}(2016){Quintero Noda}, {Shimizu}, {de la Cruz
  Rodr{\'{\i}}guez}, {Katsukawa}, {Ichimoto}, {Anan}, \&
  {Suematsu}}]{QuinteroNoda2016}
{Quintero Noda}, C., {Shimizu}, T., {de la Cruz Rodr{\'{\i}}guez}, J., {et~al.}
  2016, Mon. Not. R. Astron. Soc., 459, 3363

\bibitem[{{Scharmer} {et~al.}(2003){Scharmer}, {Bjelksjo}, {Korhonen},
  {Lindberg}, \& {Petterson}}]{Scharmer2003a}
{Scharmer}, G.~B., {Bjelksjo}, K., {Korhonen}, T.~K., {Lindberg}, B., \&
  {Petterson}, B. 2003, in \procspie, Vol. 4853, Innovative Telescopes and
  Instrumentation for Solar Astrophysics, ed. S.~L. {Keil} \& S.~V. {Avakyan},
  341--350

\bibitem[{{Scharmer} {et~al.}(2008){Scharmer}, {Narayan}, {Hillberg}, {de la
  Cruz Rodriguez}, {L{\"o}fdahl}, {Kiselman}, {S{\"u}tterlin}, {van Noort}, \&
  {Lagg}}]{Scharmer2008a}
{Scharmer}, G.~B., {Narayan}, G., {Hillberg}, T., {et~al.} 2008, \apjl, 689,
  L69

\bibitem[{{Scherrer} {et~al.}(2012){Scherrer}, {Schou}, {Bush}, {Kosovichev},
  {Bogart}, {Hoeksema}, {Liu}, {Duvall}, {Zhao}, {Title}, {Schrijver},
  {Tarbell}, \& {Tomczyk}}]{Scherrer2012}
{Scherrer}, P.~H., {Schou}, J., {Bush}, R.~I., {et~al.} 2012, \solphys, 275,
  207

\bibitem[{{Schmidt} {et~al.}(2000){Schmidt}, {Muglach}, \&
  {Kn{\"o}lker}}]{Schmidt2000b}
{Schmidt}, W., {Muglach}, K., \& {Kn{\"o}lker}, M. 2000, \apj, 544, 567,
  \dodoi{10.1086/317169}

\bibitem[{{Schmidt} {et~al.}(2012){Schmidt}, {von der L{\"u}he}, {Volkmer},
  {Denker}, {Solanki}, {Balthasar}, {Bello Gonzalez}, {Berkefeld}, {Collados},
  {Fischer}, {Halbgewachs}, {Heidecke}, {Hofmann}, {Kneer}, {Lagg}, {Nicklas},
  {Popow}, {Puschmann}, {Schmidt}, {Sigwarth}, {Sobotka}, {Soltau}, {Staude},
  {Strassmeier}, \& {Waldmann }}]{Schmidt2012}
{Schmidt}, W., {von der L{\"u}he}, O., {Volkmer}, R., {et~al.} 2012, Astron.
  Nachr., 333, 796, \dodoi{10.1002/asna.201211725}

\bibitem[{{Schou} {et~al.}(2012){Schou}, {Scherrer}, {Bush}, {Wachter},
  {Couvidat}, {Rabello-Soares}, {Bogart}, {Hoeksema}, {Liu}, {Duvall}, {Akin},
  {Allard}, {Miles}, {Rairden}, {Shine}, {Tarbell}, {Title}, {Wolfson},
  {Elmore}, {Norton}, \& {Tomczyk}}]{Schou2012}
{Schou}, J., {Scherrer}, P.~H., {Bush}, R.~I., {et~al.} 2012, \solphys, 275,
  229, \dodoi{10.1007/s11207-011-9842-2}

\bibitem[{{Shchukina} {et~al.}(2017){Shchukina}, {Sukhorukov}, \& {Trujillo
  Bueno}}]{Shchukina2017}
{Shchukina}, N.~G., {Sukhorukov}, A.~V., \& {Trujillo Bueno}, J. 2017, \aap,
  603, A98, \dodoi{10.1051/0004-6361/201630236}

\bibitem[{{Socas-Navarro} {et~al.}(2015){Socas-Navarro}, {de la Cruz
  Rodr{\'{\i}}guez}, {Asensio Ramos}, {Trujillo Bueno}, \& {Ruiz
  Cobo}}]{SocasNavarro2015}
{Socas-Navarro}, H., {de la Cruz Rodr{\'{\i}}guez}, J., {Asensio Ramos}, A.,
  {Trujillo Bueno}, J., \& {Ruiz Cobo}, B. 2015, \aap, 577, A7

\bibitem[{{Solanki} {et~al.}(2003){Solanki}, {Lagg}, {Woch}, {Krupp}, \&
  {Collados}}]{Solanki2003b}
{Solanki}, S.~K., {Lagg}, A., {Woch}, J., {Krupp}, N., \& {Collados}, M. 2003,
  \nat, 425, 692, \dodoi{10.1038/nature02035}

\bibitem[{{Spadaro} {et~al.}(2004){Spadaro}, {Billotta}, {Contarino}, {Romano},
  \& {Zuccarello}}]{Spadaro2004}
{Spadaro}, D., {Billotta}, S., {Contarino}, L., {Romano}, P., \& {Zuccarello},
  F. 2004, \aap, 425, 309, \dodoi{10.1051/0004-6361:20041004}

\bibitem[{{Su} {et~al.}(2018){Su}, {Liu}, {Li}, {Cao}, {Ahn}, \& {Ji}}]{Su2018}
{Su}, Y., {Liu}, R., {Li}, S., {et~al.} 2018, \apj, 855, 77,
  \dodoi{10.3847/1538-4357/aaac31}

\bibitem[{{van Noort} {et~al.}(2005){van Noort}, {Rouppe van der Voort}, \&
  {L{\"o}fdahl}}]{vanNoort2005}
{van Noort}, M., {Rouppe van der Voort}, L., \& {L{\"o}fdahl}, M.~G. 2005,
  \solphys, 228, 191, \dodoi{10.1007/s11207-005-5782-z}

\bibitem[{{Wiegelmann}(2004)}]{Wiegelmann2004}
{Wiegelmann}, T. 2004, \solphys, 219, 87,
  \dodoi{10.1023/B:SOLA.0000021799.39465.36}

\bibitem[{{Wiegelmann} {et~al.}(2006){Wiegelmann}, {Inhester}, \&
  {Sakurai}}]{Wiegelmann2006}
{Wiegelmann}, T., {Inhester}, B., \& {Sakurai}, T. 2006, \solphys, 233, 215,
  \dodoi{10.1007/s11207-006-2092-z}

\bibitem[{{Wiegelmann} {et~al.}(2012){Wiegelmann}, {Thalmann}, {Inhester},
  {Tadesse}, {Sun}, \& {Hoeksema}}]{Wiegelmann2012}
{Wiegelmann}, T., {Thalmann}, J.~K., {Inhester}, B., {et~al.} 2012, \solphys,
  281, 37, \dodoi{10.1007/s11207-012-9966-z}

\bibitem[{{Xu} {et~al.}(2010){Xu}, {Lagg}, \& {Solanki}}]{Xu2010}
{Xu}, Z., {Lagg}, A., \& {Solanki}, S.~K. 2010, \aap, 520, A77,
  \dodoi{10.1051/0004-6361/200913227}

\bibitem[{{Zhong} {et~al.}(2019){Zhong}, {Hou}, {Li}, {Zhang}, \&
  {Xiang}}]{Zhong2019}
{Zhong}, S., {Hou}, Y., {Li}, L., {Zhang}, J., \& {Xiang}, Y. 2019, arXiv
  e-prints, arXiv:1907.10345.
\newblock \doarXiv{1907.10345}

\bibitem[{{Zwaan}(1987)}]{Zwaan1987}
{Zwaan}, C. 1987, \araa, 25, 83, \dodoi{10.1146/annurev.aa.25.090187.000503}

\end{thebibliography}
\bibliographystyle{aasjournal}

\end{document}